# Suppressed carrier density for the patterned high mobility two-dimensional electron gas at γ-Al$_2$O$_3$/SrTiO$_3$ heterointerfaces

Wei Niu, [1,2] Yulin Gan, [1] Yu Zhang, [1] Dennis Valbjørn Christensen,[1] Merlin von Soosten, [1] Xuefeng Wang, [2,a] Yongbing Xu, [2] Rong Zhang, [2] Nini Pryds [1] and Yunzhong Chen, [1,a]

[1]*Department of Energy Conversion and Storage, Technical University of Denmark, Risø Campus, Roskilde 4000, Denmark*
[2]*National Laboratory of Solid State Microstructures, Collaborative Innovation Center of Advanced Microstructures and School of Electronic Science and Engineering, Nanjing University, Nanjing 210093, China*
[a]*Author to whom correspondence should be addressed. Electronic mail: yunc@dtu.dk; xfwang@nju.edu.cn*

**Abstract:**

The two-dimensional electron gas (2DEG) at the non-isostructural interface between spinel γ-Al$_2$O$_3$ and perovskite SrTiO$_3$ is featured by a record electron mobility among complex oxide interfaces in addition to a high carrier density up to the order of $10^{15}$ cm$^{-2}$. Herein, we report on the patterning of 2DEG at the γ-Al$_2$O$_3$/SrTiO$_3$ interface grown at 650 °C by pulsed laser deposition using a hard mask of LaMnO$_3$. The patterned 2DEG exhibits a critical thickness of 2 unit cells γ-Al$_2$O$_3$ for the occurrence of interface conductivity, similar to the unpatterned sample. However, its maximum carrier density is found to be approximately $3\times10^{13}$ cm$^{-2}$, much lower than that of the unpatterned sample ($\sim10^{15}$ cm$^{-2}$). Remarkably, a high electron mobility of approximately 3,600 cm$^2$V$^{-1}$s$^{-1}$ was obtained at low temperatures for the patterned 2DEG at a carrier density of $\sim 7\times10^{12}$ cm$^{-2}$, which exhibits clear Shubnikov-de Hass quantum oscillations. The patterned high-mobility 2DEG at the γ-Al$_2$O$_3$/SrTiO$_3$ interface paves the way for the design and application of spinel/perovskite interfaces for high-mobility all-oxide electronic devices.

Two-dimensional electron gases (2DEGs) formed at SrTiO$_3$-based interfaces provide a rich platform for fundamental research and device applications[1]. Their unique properties, such as superconductivity[2], magnetism[3], high carrier mobility[4] and sensitivity to light illumination[5], have drawn extensive interests. Among complex oxide interfaces, the isostructural perovskite-type LaAlO$_3$/SrTiO$_3$ (LAO/STO) interface is so far the most investigated system. Nevertheless, although extensive research has been carried out on this system, the typical mobility remains ~ 1,000 cm$^2$V$^{-1}$s$^{-1}$ or less (at low temperatures). Recently, a new 2DEG was discovered at the non-isostructural interface between perovskite STO and spinel γ-Al$_2$O$_3$ (GAO) with compatible oxygen sublattices[6-10]. Remarkably, the GAO/STO heterostructure shows much higher electron mobility (greater than 140, 000 cm$^2$V$^{-1}$s$^{-1}$) as well as extremely high carrier densities of more than 10$^{15}$ cm$^{-2}$.[6] Moreover, micro-patterning of complex oxides with conventional semiconductor techniques is highly needed to meet the promise for post-silicon electronics, i.e. to integrate complex oxides interfaces into integrated chips and spintronics devices. Although nano-patterned interfaces by conducting-atomic force microscopy (c-AFM)[11] have been demonstrated, micro-patterning of complex oxides has been proven to be challenging. This so far has been implemented primarily using amorphous LaAlO$_3$- or AlO$_x$- hard masks [12-15] or Ar-ion beam irradiation[16]. These processes, generally, require additional care as the deposition of amorphous LAO or AlO$_x$ layers or the Ar-ion irradiation can by itself induce conductivity in STO,[7,17,18] leading to failure of the patterned devices. In contrast to the chemically active hard masks or irradiation, a chemically inert mask of manganites (which shows little redox reaction with STO[16]) has also been applied to pattern oxide interfaces, particularly the 2DEG in a-LAO/STO system formed at room temperature[19]. Whether this technique can also be applicable to pattern the 2DEG grown at high-temperatures, where significant oxygen exchange and cation intermixing across the interface could occur, has yet to be investigated.

In this letter, we present the high temperature patterning of the 2DEG at the GAO/STO interface with LaMnO$_3$ as a hard mask. The high-mobility 2DEG is conserved in the patterned structures, but a much suppressed carrier density was obtained, probably due to the presence of the manganite hard mask. Moreover, clear quantum oscillations were observed at these patterned spinel/perovskite interfaces. The balance between the high mobility and low carrier density in patterned GAO/STO interfaces is a step forward to integrate high quality oxide interfaces in future devices.

The Hall bar devices were fabricated by initially depositing an amorphous LaMnO$_3$ (a-LMO) layer (50 nm) (see Fig. 1) on TiO$_2$-terminated STO (001) substrates[20,21] using pulsed laser deposition (PLD) at room temperature. The a-LMO/STO heterointerface was found to be insulating regardless of the deposition oxygen pressure. Optical lithography was then used to create patterned structures with microscale dimension. Subsequently, the exposed a-LMO was removed by selective wet chemical etching so that the bare STO is patterned in a Hall bar geometry[22,23]. After removing the residual photoresist with a lift-off procedure, the patterned substrate was transferred into the PLD chamber for the deposition of GAO. The growth of GAO was performed at 650 ºC with an oxygen pressure of 1×10$^{-5}$ mbar, and the samples were cooled down at the growth oxygen pressure at a rate of 15 ºC/min to room temperature after deposition. For comparison, unpatterned 5×5 mm$^2$ GAO/STO reference samples were prepared under the same growth conditions and measured in the van der Pauw geometry. For transport measurements in both Hall-bar and van der Pauw geometry, ultrasonically wire-bonded aluminum wires were used as electrodes. For the patterned sample, the film thickness, $t$, was controlled by the growth rate, which was determined with the unpatterned sample by reflection high-energy electron diffraction (RHEED) oscillations[6]. By carefully optimizing the film growth conditions, $t$ can be controlled down to a quarter of the unit cell (uc), i.e. $a/4$~0.2 nm [6,24].

Figure 2(a) shows an optical micrograph of a typical patterned device where the width of the Hall bar is 50 μm and length between two voltage probes is 500 μm. A six-probe configuration of the Hall bar allows for the measurement of both longitudinal and Hall resistance at the same time. Figure 2(b) shows the temperature-dependent sheet resistances of GAO/STO Hall bar devices. The interfacial conduction depends critically on the thickness of GAO film. When the thickness of GAO is thinner than 1.75 uc, the interface is highly insulating. At $t$=1.75 uc, the sample becomes metallic but shows carrier freezing out at $T$≤100 K. For $t$≥2 uc, the interfaces show metallic behaviors over the whole temperature regime down to the base temperature of 2 K. As shown in Fig. 2(c) and (d), the corresponding temperature-dependent sheet carrier density, $n_s$, and mobility, $\mu$, are deduced from the measurements of the linear Hall coefficient $R_H$, using $n_s=-1/R_H e$. The carrier density seems to be separated into two groups, samples with similar carrier density in the range of 2-2.5 uc (7-8×10$^{12}$ cm$^{-2}$) and samples between 2.5-10 uc (2-3×10$^{13}$ cm$^{-2}$), see Fig. 2(c). The highest mobility of 3,600 cm$^2$V$^{-1}$s$^{-1}$ at 2 K was obtained for $t$=2.25 uc ($n_s$~ 7×10$^{12}$ cm$^{-2}$). Additionally, the pattered 2DEGs with $\mu$>1000 cm$^2$V$^{-1}$s$^{-1}$ at 2K are only detected in the thickness range of 2 uc ≤ t ≤ 2.5 uc. This thickness range of high-mobility is comparable to the 2 uc ≤ t ≤ 3uc observed for unpatterned samples[6].

Figures 3(a) and (b) summarize the thickness dependent sheet conductance ($\sigma_s$) and the carrier density ($n_s$) respectively, of the patterned samples measured at room temperature. When $t$ is increased from 1 uc to 2 uc, the $\sigma_s$ and $n_s$ of interfaces jump more than 4 orders, accompanied with the sharp transition from insulating to metallic state. This critical thickness behavior of Hall bar interfaces is in good agreement with the unpatterned GAO/STO interface[6]. However, as illustrated in Fig. 3(b), the carrier density ($n_s$) of the pattered samples is always in the range of 0.7- 3×10$^{13}$ cm$^{-2}$, although they were deposited at an oxygen pressure of 10$^{-5}$ mbar. This is dramatically different from the unbuffered sample, where a peak carrier density up to 1×10$^{15}$ cm$^{-2}$ is obtained in the range of 2 uc ≤ $t$ ≤ 3uc[6]. Notably, the critical thickness dependence of the carrier density for both the patterned and unpatterned samples are highly reproducible. The suppression of the carrier density in pattered samples is most likely due to the presence of the manganite buffer layer. This is because the GAO/STO heterostructure is one of the typical STO-based heterostructures, where the interface conductivity originates mainly from oxygen vacancies due to interfacial redox reactions[6,7,9,10]. At high deposition temperatures, the oxygen ions in STO can diffuse over many micrometers in minutes[25]. Therefore, a significant transfer of oxygen from STO to GAO is expected. This could account for the high concentration of oxygen vacancies as well as high density of charge carriers at the interface of the unpatterned GAO/STO. Different from the GAO film which can promote the formation of oxygen vacancies in STO by chemical redox reactions[17], the LMO film is one of the most outstanding oxides which show no degradation of the STO. This is due to the fact that the bottom of the LMO conduction band is about 1 eV lower than that of STO, any reduction, if occurs, is preferably on the LMO side, i.e. the reconstructed electrons will be firstly transferred to the Mn sublattice before filling the electronic shell of Ti ions[4,26]. Moreover, LMO could activate the oxygen uptaking in STO due to its catalytic activity for oxygen reduction reaction at high temperatures[27]. In this context, much less oxygen vacancies are expected in the patterned GAO/STO thus the suppressed carrier density. Additionally, the change in the profile of oxygen vacancies could also account for the difference in mobility between patterned and unpatterned samples.

Finally, the high mobility of our patterned 2DEG together with the low carrier density is further confirmed by experimental observation of Shubnikov-de Hass (SdH) oscillations[6,28,29]. Figure 4(a) shows the longitudinal resistance of the $t$=2.25 uc sample. The magnetic field (up to 16 T) was applied perpendicular to the interface at $T$=2 K. The oscillations superimposed on a positive background are visible directly in the raw magnetoresistance data for magnetic fields larger than 6 T. After removing a smooth background, the magnetoresistance exhibits oscillations presented in Fig. 4(b), which are periodic with $1/B$. The inset of Fig. 4(b) shows the position of the oscillation peak in $1/B$ versus the effective Landau level. The fitted line (blue dash line) indicates the SdH frequency of $F$= 71.8T. The carrier density can be estimated from SdH oscillation by the formula: $n_{2D} = g_V g_S eF/h$, where $g_V$, $g_S$ and $F$ are the valley degeneracy, spin degeneracy and SdH frequency, respectively. By taking a single valley and $g_S$=2,[6] the carrier density was calculated to be $n_{2D}$= 3.47×10$^{12}$ cm$^{-2}$. Notably, this carrier density deduced by the SdH oscillation is slightly lower than that obtained from the Hall effect (7×10$^{12}$ cm$^{-2}$), which is common for 2DEG in STO-based heterointerfaces, such as those at the LAO/STO interface[28,29] and in La- or Nb-doped STO heterostructures[30,31]. This discrepancy is either due to the fact that a fraction of carriers measured bythe Hall effect do not satisfy the conditions for the SdH oscillation[6,28-30,32], or due to the presence of multiple quantum wells[29,33].

In conclusion, we have demonstrated the fabrication of patterned 2DEG at GAO/STO interfaces with high mobility using a manganite hard mask. Compared with unpatterned GAO/STO heterostructures analogues, suppressed carrier density is obtained in the patterned interface. The relatively high electron mobility and low carrier density enables the study of quantum oscillations at GAO/STO interfaces. This patterning method provides not only the possibility of making patterned interface devices with high mobility, but also a step forward to integrate high quality spinel/perovskite oxide interfaces for device applications.

We thank Jørgen Stubager for the technical assistance. Wei Niu thanks the support by China Scholarship Council. X.F.W. acknowledges the financial support from the National Key Projects for Basic Research of China under Grant No. 2014CB921103 and the Collaborative Innovation Center of Solid-State Lighting and Energy-Saving Electronics.

**Figure Caption**

FIG. 1 Schematic illustration of the patterning process for a GAO/STO Hall bar device. The conductive interface of 2DEG is only formed at the interface between GAO and STO, which is illustrated by the red layer in the cross-section.

FIG. 2. (a) Optical microscopy image of the Hall bar device with a channel width of 50 μm and a distance between longitudinal voltage probes 500 μm apart. (b) Temperature dependence of sheet resistance for the interface conduction at different GAO thicknesses. (c), (d) Temperature dependence of carrier density ($n_s$) and electron Hall mobility ($\mu$), respectively, for the interface conduction at the different GAO thicknesses.

FIG. 3. (a) Thickness-dependent sheet conductance measured at 300K. (b) Comparison of thickness-dependent carrier density between patterned Hall bar devices and unpatterned van der Pauw devices.

FIG. 4 Shubnikov-de Hass oscillations of the conduction at GAO/STO interface. (a) Longitudinal resistance, $R_{xx}$, as a function of magnetic field with SdH oscillations at 2 K for the 2.25 uc sample. (b) Amplitude of the SdH oscillation versus the reciprocal magnetic field. The inset shows the index plots of 1/$B$ versus the effective Landau level.

**Figures:**

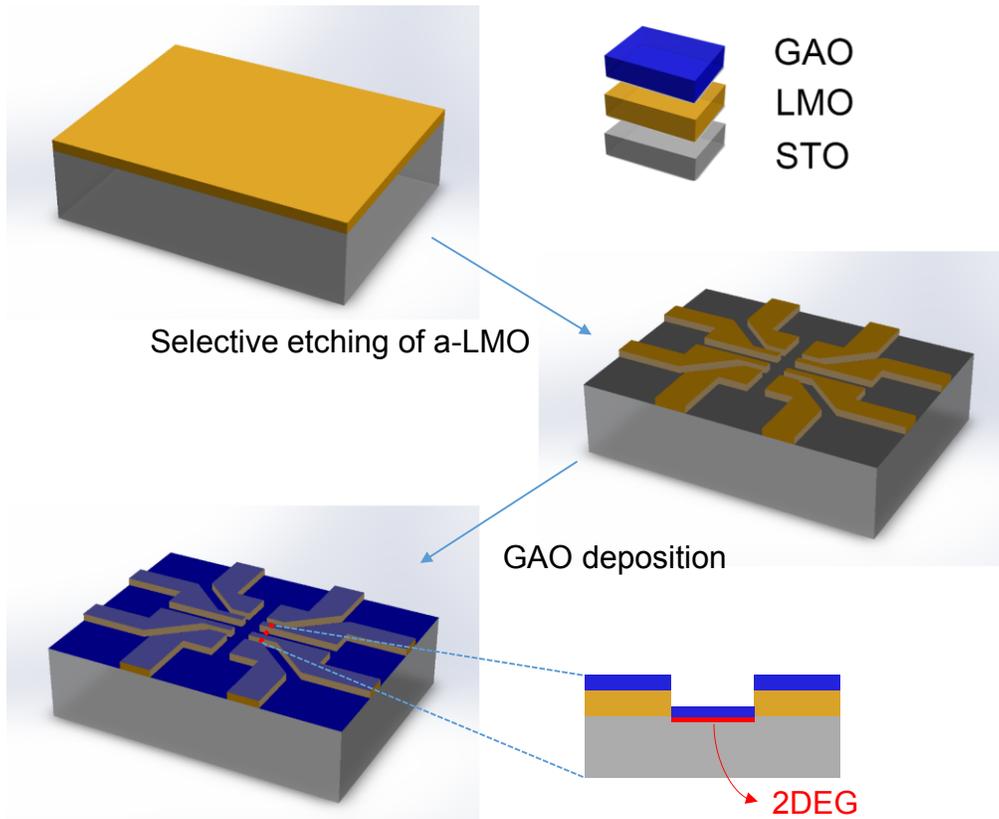

**FIG. 1** Schematic illustration of the patterning process for a GAO/STO Hall bar device. The conductive interface of 2DEG is only formed at the interface between GAO and STO, which is illustrated by the red layer in the cross-section.

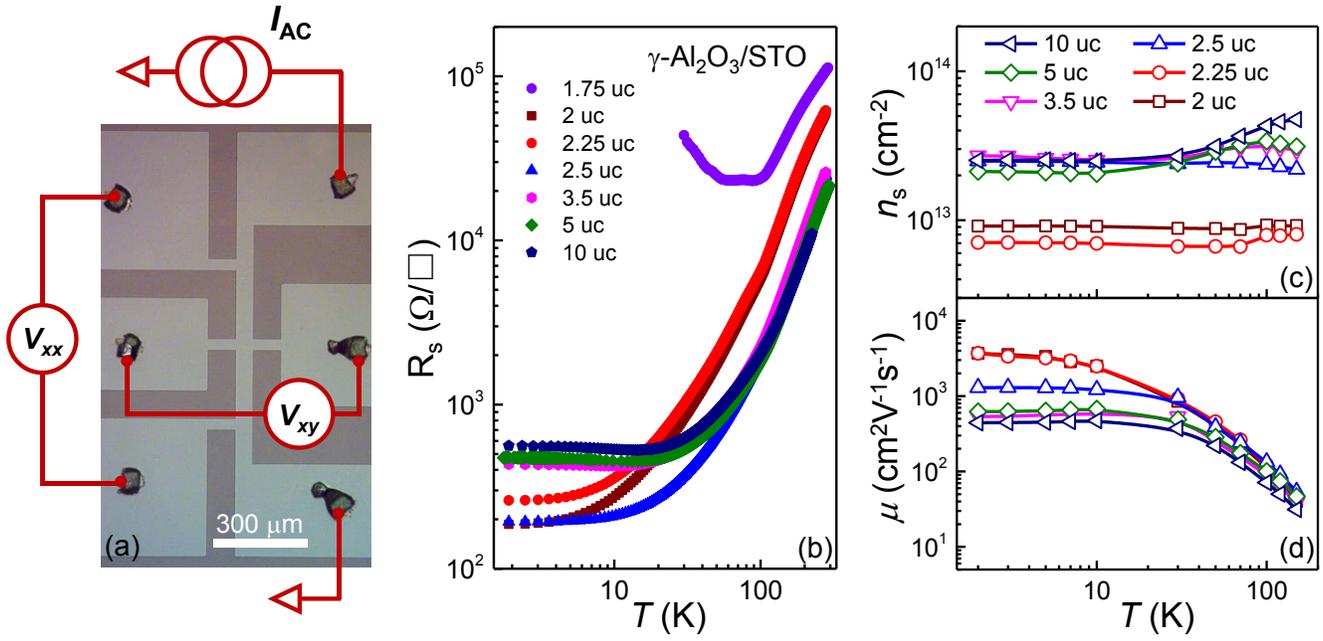

**FIG. 2**. (a) Optical microscopy image of the Hall bar device with a channel width of 50 μm and a distance between longitudinal voltage probes 500 μm apart. (b) Temperature dependence of sheet resistance for the interface conduction at different GAO thickness. (c), (d) Temperature dependence of carrier density ($n_s$) and electron Hall mobility ($\mu$) for the interface conduction at different GAO thickness.

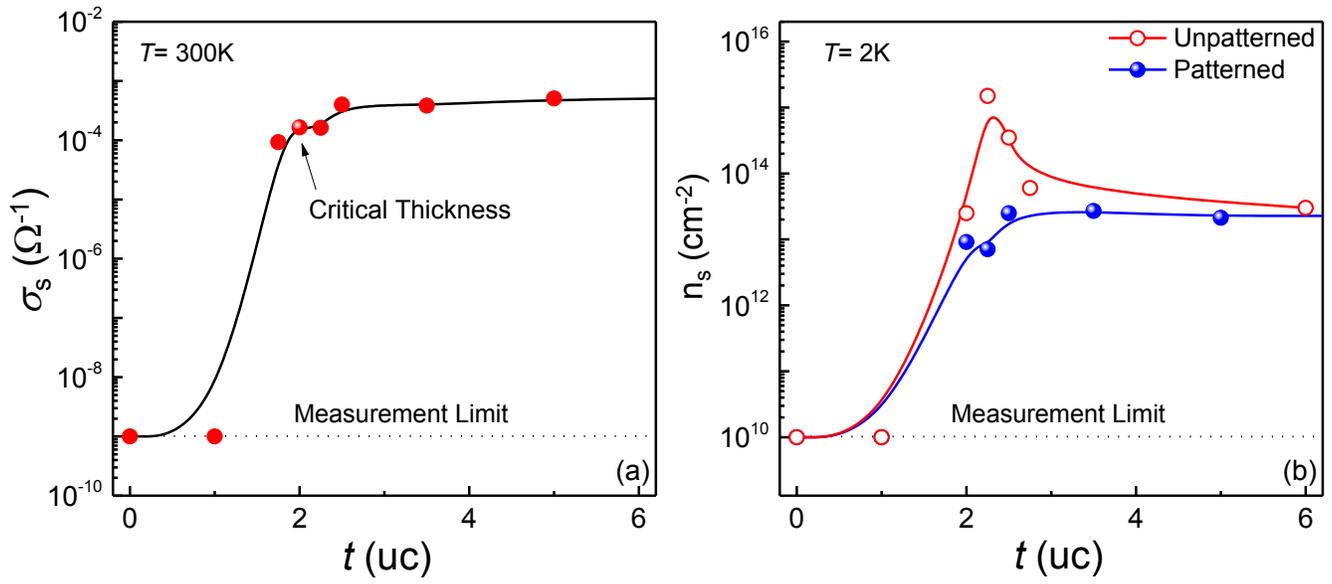

**FIG. 3.** (a) Thickness-dependent sheet conductance measured at 300K. (b) Comparison of thickness-dependent carrier density between patterned Hall bar devices and unpatterned van der Pauw devices.

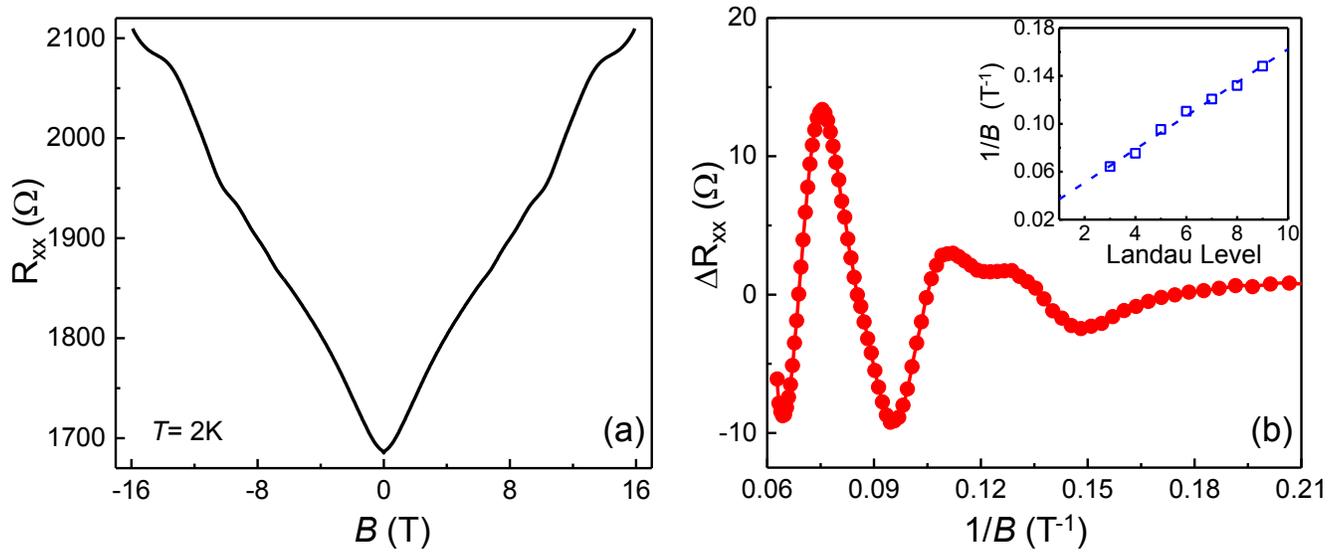

**FIG. 4 Shubnikov-de Hass oscillation of the conduction at GAO/STO interface.** (a) Longitudinal resistance, $R_{xx}$, as a function of magnetic field with SdH oscillations at 2K for the 2.25 uc sample. (b) Amplitude of the SdH oscillation versus the reciprocal magnetic field. The inset shows the position of the oscillation peak in 1/B versus the effective Landau level.